\documentclass[fleqn,usenatbib,useAMS]{mnras}

\usepackage{newtxtext,newtxmath}
\usepackage[T1]{fontenc}
\usepackage{ae,aecompl}

\usepackage{graphicx}	% Including figure files
\usepackage{amsmath}	% Advanced maths commands
\usepackage{amssymb}	% Extra maths symbols
\usepackage{multicol}
\usepackage{bm}		% Bold maths symbols, including upright Greek
\usepackage{pdflscape}	% Landscape pages
\usepackage{hyperref}
\usepackage[noabbrev]{cleveref}   % automatic reference manager
\title[The Solar Wind In Time]{The solar wind in time: a change in the behaviour of older winds?}

\author[D. \'{O} Fionnag\'{a}in \& A. A. Vidotto]{
D. \'{O} Fionnag\'{a}in\thanks{Email: ofionnad@tcd.ie}
and A. A. Vidotto
\\
School of Physics, Trinity College Dublin, College Green, Dublin 2, Ireland\\
}

\date{Accepted XXX. Received YYY; in original form ZZZ}

\pubyear{2017}

\begin{document}
\label{firstpage}
\pagerange{\pageref{firstpage}--\pageref{lastpage}}
\maketitle

\begin{abstract}
In the present paper, we model the wind of solar analogues at different ages to investigate the evolution of the solar wind. Recently, it has been suggested that winds of solar type stars might undergo a change in properties at old ages, whereby stars older than the Sun would be less efficient in carrying away angular momentum than what was traditionally believed. Adding to this, recent observations suggest that old solar-type stars show a break in coronal properties, with a steeper decay in X-ray luminosities and temperatures at older ages. We use these X-ray observations to constrain the thermal acceleration of winds of solar analogues. Our sample is based on the stars from the `Sun in time' project with ages between 120-7000 Myr. The break in X-ray properties leads to a break in wind mass-loss rates ($\dot{M}$) at roughly 2 Gyr, with $\dot{M}$ (t < 2 Gyr) $\propto t^{-0.74}$ and $\dot{M}$ (t > 2 Gyr) $\propto$ $t^{-3.9}$. This steep decay in $\dot{M}$ at older ages could be the reason why older stars are less efficient at carrying away angular momentum, which would explain the anomalously rapid rotation observed in older stars. We also show that none of the stars in our sample would have winds dense enough to produce thermal emission above 1-2 GHz, explaining why their radio emissions have not yet been detected. Combining our models with dynamo evolution models for the magnetic field of the Earth we find that, at early ages ($\approx$100 Myr) our Earth had a magnetosphere that was 3 or more times smaller than its current size.
\end{abstract}

\begin{keywords}
stars: mass-loss -- stars: solar-type -- stars: winds, outflows
\end{keywords}

%%%%%%%%%%%%%%%%%%%%%%%%%%%%%%%%%%%%%%%%%%%%%%%%%%

%%%%%%%%%%%%%%%%% BODY OF PAPER %%%%%%%%%%%%%%%%%%

\section{Introduction}
\label{sec:intro}
Solar analogue stars lose angular momentum and mass through stellar winds. These magnetised winds determine how the rotation of a star will decay with time \citep{weberdavis67, vidotto11}, although the exact processes behind this occurrence is not fully understood. These winds are assumed to be homologous to the solar wind at different ages, which is composed of a fully ionised plasma that streams outwards from the Sun \citep{parker58}. As these stars similar to our own Sun age, various properties seem to evolve over time, such as rotation and magnetic activity \citep{skumanich72,dorren94,ribas05,guinan09,vidotto14}. Since the magnetic activity of a star is one of the preeminent factors determining how active the stellar wind is \citep{wood02}, it suggests that the stellar wind will also trend in a similar fashion. 

	\subsection{Evolution of stellar activity}
    \label{sec:stellar-activity}
    Stellar magnetic activity is observed to decline with age and rotation \citep{skumanich72,ribas05,vidotto14}. There are many proxies for the activity of a star such as rotation, magnetism, chromospheric emission  and X-ray luminosity. In recent years, there has been a surge in observation-based research suggesting a break in solar analogue activity as stars cross a certain rotation or age threshold.
    \citet{vansaders16} modeled a set of 21 older stars that have been observed by Kepler and reported the abnormally rapid rotation in older main sequence (MS) stars, which does not agree with previous period-age relations. They suggested that magnetic braking seems to weaken significantly in evolved MS stars at Rossby number\footnote{$Ro$ represents Rossby number, which is the ratio between stellar rotation and convective turnover time. $Ro = P_{\rm rot}/\tau_{\rm conv}$} $Ro \approx$ 2, which they assumed corresponds to when the stars reach an age of $\approx$ 2-4 Gyr. Since the rate of angular momentum loss is related to the mass-loss rate \citep{weberdavis67,vidotto14}, this break in angular momentum loss is likely to be associated with a decline in mass-loss rate ($\dot{M}$). We investigate this further in the present work.
    Recently, \citet{booth17} have shown that X-ray luminosity declines more rapidly for stars older than $\approx$1 Gyr. They found a steep decrease in the age-activity relationship and suggested this could be due to increased stellar spin-down. Their explanation, however, contradicts the findings of \citet{vansaders16} which found unusually high rotation rates in older stars. Here, we present a way to simultaneously explain the observations from Booth et al. 2017 and van Saders et al. 2016; our suggestion is that the observed decrease in X-ray luminosity is linked to a weaker stellar wind, which removes less angular momentum and thus, allows for higher rotation rates in older stars, as seen in Kepler observations. 
    \citet{metcalfe16} examined chromospheric activity observations from calcium lines of solar analogues. They suggested that the break in activity is caused by a change in the dynamo properties at approximately the solar age which is related to the break observed by \citet{vansaders16}. \citet{kitchatinov17} recently demonstrated that by switching off the global dynamo once a critical rotation period is reached, a similar decline in stellar spin down and magnetism can be achieved for older stars.  
    \citet{vidotto14} (Figure 2 within) showed that there is a trend in surface averaged magnetic field with age for solar-type stars, a power law dependence was found but there is an apparent drop in magnetic activity for stars older than $\approx$2-2.5 Gyr. 
    Another break in behaviour in solar-type stars is presented by \citet{beck17} (Figure 5 within). They have shown that there is a break in lithium abundances in solar analogue stars which drift beneath a surface rotation velocity of $\approx$ 2-3 km/s. Each of these works suggests that there exists a transition between regimes for aged low-mass stars past 1 Gyr. While the nature of this transition has not yet been entirely defined, there is enough evidence for further investigation.\par
\begin{figure}
    \centering
    \includegraphics[width=\hsize]{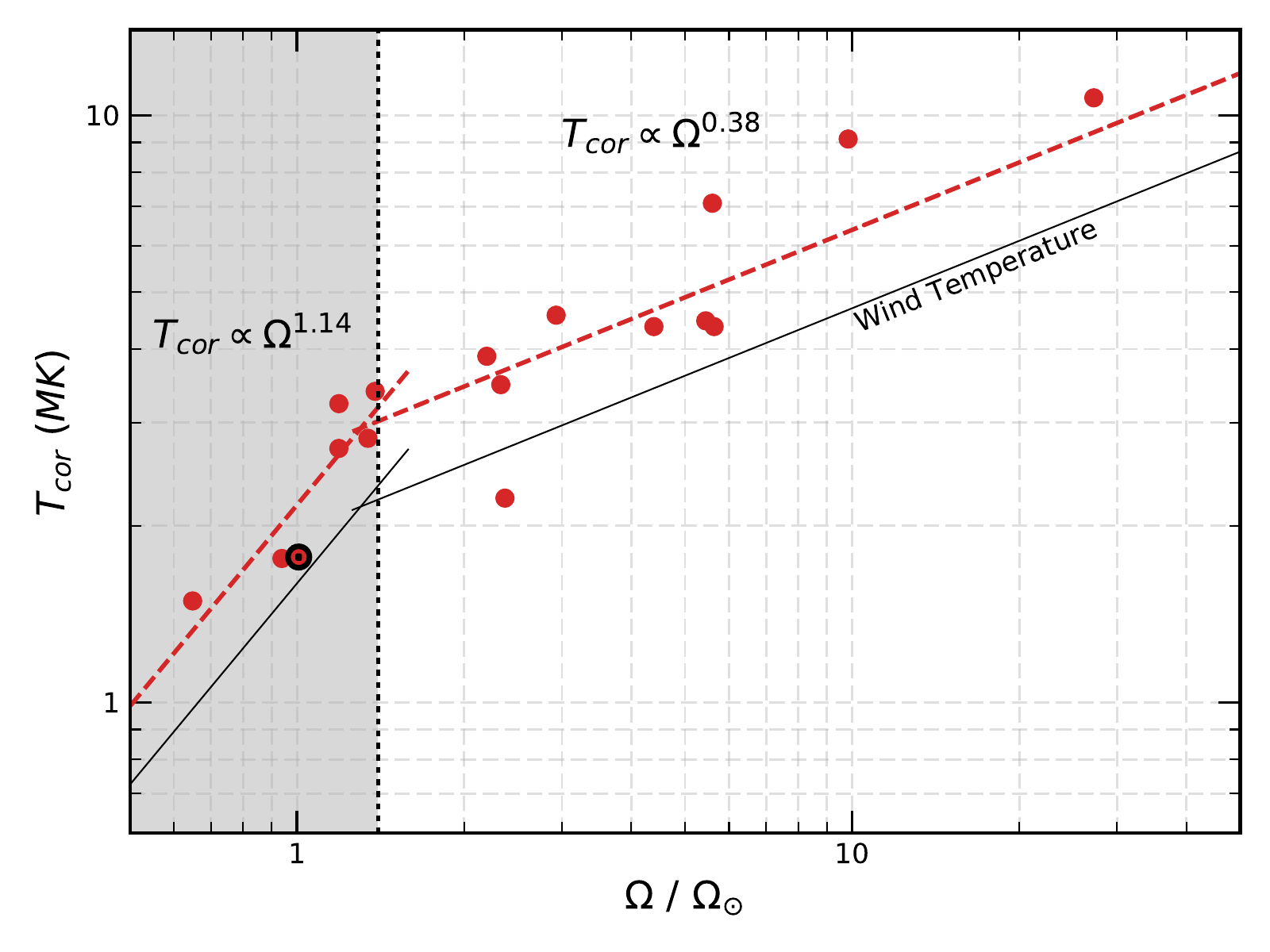}
      \caption{Average coronal temperature ($T_{cor}$) values derived from X-ray observations \citep{johnstone15c} show a strong correlation with rotation (broken red line). This relation was scaled down to give a base wind temperature, which we use for our simulations. This break in temperatures occurs at 1.4 $\Omega_{\odot}$, and while the physical mechanism for this break is not understood, it would imply some transition between regimes for solar-type stars. Interesting to note is that the Sun ($\odot$, representing solar average in the solar cycle) has just evolved past this transition. Both fits are shown in \Cref{eq:Tcorlow,eq:Tcorhigh,eq:rot-T,eq:rot-T2}. The statistical significance of these fits are discussed further in \cref{app:A}.
              }
         \label{fig:tcor}
   \end{figure}

 
    Thermally-driven winds are effected by the temperature at their base, with higher temperatures leading to faster winds. Currently, defining the temperature at the base of the wind of solar-type stars is not possible through observations and we must rely on empirical methods. \citet{johnstone15c} took coronal temperatures of low-mass main sequence stars and showed how they they are correlated to X-ray surface fluxes \citep{telleschi05}. 
    Here, we find additional evidence that the coronal temperatures of solar-like stars show a steeper decay for older, slowly rotating stars (\Cref{fig:tcor}). Coronal temperatures are from \citet{johnstone15c}, with rotation rates taken from \citet{raassen03,telleschi05,wood06,wood10,gudel07,vidotto14}. We have excluded M dwarfs from their sample, so as to limit the trend found to solar-type stars. \Cref{fig:tcor} shows that there is an evident break in coronal temperature at $\approx$ 1.4 $\Omega_{\odot}$. This break in behaviour at lower rotation rates results in power laws over two different regimes
    \begin{equation}
    \label{eq:Tcorlow}
    T_{\rm cor}\ (\Omega < 1.4\ \Omega_{\odot})\ \propto\ \Omega^{1.20}
    \end{equation}
    \begin{equation}
    \label{eq:Tcorhigh}
    T_{\rm cor}\ (\Omega > 1.4\ \Omega_{\odot})\ \propto\ \Omega^{0.37} .    
    \end{equation} 
    The $\Omega \approx 1.4\ \Omega_{\odot}$ break occurs at $\approx$2 Gyr for the sample of stars used, which is around the same age as those found by \citet{booth17} ($\approx$1 Gyr) and not dissimilar to ages found by \citet{vansaders16} ($\approx$ 2-4 Gyr). Although these values are not identical they are a good match considering limitations in age constraints on these stars. From convective turnover times for solar mass stars \citep{kiraga07}, we find this break occurring at $Ro = 1.14$.
     Note that, the break in behaviour here is inherent to older solar-type stars above $\approx$ 1-2 Gyr. This argument does not preclude the existence of a suggested sudden change in rotational braking at young ages, e.g. \citet{johnstone15a} and \citet{gondoin17}.
    
 	\subsection{The Sun in time project}
 	The `Sun in Time' project is used as the basis for selecting our sample for our study, discussed further in \Cref{sec:solarwindsample}. The project was created to explore the life-long activity evolution of our own Sun, from when it reached the main sequence, by studying a group of solar-mass stars (e.g. \citealt{gudel07}).  \citet{dorren94} looked at the optical and UV Sun in time, from which they could infer declining trends in activity with age. \citet{gudel97} examined how the high energy radiation (X-ray and ultraviolet) emitted from the Sun has evolved over time. They used X-ray observations of nine solar-like G type stars to probe their coronae and used this as a proxy for an evolving Sun. Since these high energy fluxes can be connected to coronal activity, they derived trends in coronal temperature and emission measure with rotation and age of these solar analogues (see also \citealt{telleschi05}). \citet{ribas05} built on this previous work and investigated how these high irrandiances would effect planetary atmospheres. The Sun in time project inspired our investigation on how the solar wind has evolved during the main sequence. 
    Understanding how the wind of the Sun has evolved is an important step in understanding the long-term evolution of the planets in the solar system, including the Earth and the development of life (see e.g. \citealt{chassefiere04}).

    \subsection{Planetary environments}
    Stellar winds originate at the stellar surface and interact with all bodies in their path as they expand into the interplanetary medium, up to the astropause. Therefore, similarly to planets in our solar system, exoplanets are expected to react to changes in the wind. In the solar system, the Earth has the largest magnetosphere of the terrestrial planets. This magnetosphere shields the Earth from the harmful solar wind and also dynamically changes with the evolving solar wind \citep{cravens97,bagenal13} and planetary dynamo. If the wind of the Sun changes on evolutionary timescales, then it is expected that the Earth's magnetosphere will evolve on similar timescales. In practicality this will also be effected on timescales related to the changing internal planetary dynamo \citep{zuluaga13}. This will have significant effects along the evolution of the Earth regarding retention of its atmosphere. 
    Any erosion of atmosphere would also have significant implications for the development of any life on a planet \citep{chassefiere04}. Mars possesses an induced magnetosphere, which does not originate from the interior of the planet, but rather the build up of ions near the ionosphere where the wind impacts the upper atmosphere producing currents and their own magnetic field \citep{bertucci11}.
    
    The goal of this paper is to construct an overall picture of solar wind evolution by modeling winds of solar-analogues. Our models predict stellar mass-loss rates, wind radio emission and can be used to predict how the local environment around the Earth evolves with time. 
    In Section 2 we define the model used for these winds and the input parameters required. We describe the sample of stars used, and how we calculate radio emission from each of their winds. Section 3 describes the results we find for global wind properties. This includes mass-loss rates and radio emission from the winds and local wind properties, which focuses on the local conditions around a fictitious Earth orbiting each of the stars in our sample. 
In Section 4 we include a discussion on the results and their significance compared to other works. We finalise by drawing conclusions from this work and summarising in Section 5.

\section{Stellar Wind Modeling}
\label{sec:polytropic}

	The first work to propose that stars undergo mass loss was \citet{parker58}. He suggested that the solar wind is in hydrodynamical expansion. However, the isothermal nature of Parker winds leads to excessive acceleration of the wind as it expands radially, leading to exaggerated wind velocities and temperatures at distances far from the solar surface. Polytropic models alleviate this issue as they do not constrain the wind to being isothermal, which allows a more reasonable representation of the wind at further distances \citep{keppens99}. Here, we present one-dimensional (1D) thermally-driven hydrodynamic wind model that is used to compute the steady state solutions of the winds of stars from the `sun in time' sample (\Cref{table:sample}). We do not explicitly include magnetism in the wind equations, but we note that its presence is implicitly assumed as the cause of the MK temperatures of the winds. Polytropic winds follow the momentum equation
     \begin{equation}
     \label{eq:polytropic}
     v \frac{dv}{dr} + \frac{1}{\rho} \frac{dp}{dr} + \frac{GM_{\star}}{r^2} = 0 , 
     \end{equation}
where $v$ is the velocity of the wind, $\rho$ is the mass density of the wind, $p$ represents the pressure of the wind, $M_{\star}$ is the stellar mass, $G$ is the gravitational constant and $r$ represents distance from the stellar centre. The first term in \Cref{eq:polytropic} represents the acceleration of the wind, which is produced by the pressure gradient  and gravity (second and third terms respectively). Since the wind in this case is polytropic, the temperature and pressure change with density
     \begin{equation}
     \centering
     \label{eq:polytemprho}
     T = T_{0} \left( \frac{\rho}{\rho_{0}} \right)^{\Gamma - 1}, \hspace{2cm} p = p_{0} \left( \frac{\rho}{\rho_{0}} \right)^{\Gamma} ,
     \end{equation}
	where \(\Gamma\) is the polytropic index, and represents the energy deposition in the wind (when \(\Gamma = 1\) the wind is isothermal). \(T_{0}\), \(p_{0}\) and \(\rho_{0} \) represent the base temperature, pressure and density of the wind respectively. Methods of defining base temperature and density are discussed in \Cref{sec:temp-rot,sec:den-rot}.
     Our 1D wind model assumes a spherically symmetric, steady wind, that behaves similarly to the Parker wind solution \citep{parker58}, except the energy deposition in the wind is altered to be less than that of an isothermal wind. This change in energy deposition slows the expansion of the wind as $\Gamma$ is increased, giving rise to slower, denser winds. It begins with a subsonic flow, which transitions to a supersonic flow once it passes through the critical radius (known as the sonic point): 
     \begin{equation}
     r_c = \frac{G M_{\star}}{2 \Gamma [a(r_c)]^2} ,
     \end{equation}     
     where $a(r_c)$ is the sound speed at the sonic point.      
     To benchmark our simulations of the `solar wind in time', we constrain the parameters of our model so as to best reproduce the solar wind properties. The Sun is the only star for which we have direct wind measurements. In the solar wind, \citet{vandoor11} derived an effective polytropic index of $\Gamma$=1.1. Numerical models of solar-type stars usually adopt a range of 1.05 to 1.15 for $\Gamma$ \citep{matt12,johnstone15a,johnstone15b,vidotto15,keppens99}. In our model, we adopt a value of 1.05. 
     To further reproduce observations of the solar wind, we adopt a base wind density of $2.2\times10^8$ cm$^{-3}$, which is consistent with observations of coronal hole densities \citep{warren09}. We use a wind base temperature of 1.5 MK, which, in conjunction with our $\Gamma$ = 1.05, reproduces the solar wind velocities observed at the Earth, $v_{\oplus} = 443$ kms$^{-1}$, which is consistent with observations (\citealt{mccomas08,usmanov14}). Our model predicts a number density of $10.5 cm^{-3}$ at the Earth's orbit, which is also consistent with observations \citep{murdin00,bagenal13,usmanov14}. At the martian orbital distance, we find a wind density and velocity of 12 cm$^{-3}$ and 450 km/s. These model values agree with observations made by the MAVEN spacecraft \citep{lee17}. Finally, our model predicts a solar wind mass-loss rate of $3.5\times10^{-14} M_{\odot}$ yr$^{-1}$, which reproduces the observed values presented in \citet{wang98}.
    
    \subsection{Solar wind in time sample}
	\label{sec:solarwindsample}
\begin{table}
\caption{Sample of stars used in the present study: This sample is similar to that used in The Sun In Time sample \citep{gudel97,gudel07}, with the omission of $\beta$ Hyi and 47 Cas B. Values are mostly taken from \citet{gudel07,vidotto14}. The X-ray luminosity of the sun here is considered to be between maximum and minimum. Errors in age shown can be found in a{)} \citet{stauffer98}, b{)} \citet{lopez06}, c{)} \citet{king03}, d{)} \citet{perryman98}, e{)} \citet{mamajek08}, f{)} \citet{liu02}, g{)} \citet{ramirez14}, h{)} \citet{metcalfe15}.}             % 	title of Table
\begin{minipage}{0.47\textwidth}
\label{table:sample}      % is used to refer this table in the text
\centering                          % used for centering table
\resizebox{\columnwidth}{!}{%
%\resizebox{\textwidth}{!}{
\begin{tabular}{l c c c c c c}        % centered columns (4 columns)
\hline\hline                 % inserts double horizontal lines
Star & M & R & P$_{\rm rot}$ & Age & $log[L_X]$ & d \\    % table heading 	
     & (M$_{\odot}$) & (R$_{\odot}$) & (d) & (Gyr) & (erg/s) & (pc)\\
\hline                        % inserts single 	horizontal line
   EK Dra & 1.04 & 0.97 & 2.77 & 0.12$\pm$0.008$^a$ & 29.93 & 34.5\\	
   HN Peg & 1.10 & 1.04 & 4.55 & 0.26$\pm$0.046$^b$ & 29.00 & 17.95\\
   $\chi^1$ Ori & 1.03 & 1.05 & 4.83 & 0.5$\pm$0.1$^c$ & 28.99 & 186.0\\
   $\pi^1$ UMa & 1.00 & 1.00 & 5 & 0.5$\pm$0.1$^c$ & 28.97 & 14.36\\
   BE Cet & 1.09 & 1.00 & 12.4 & 0.6$\pm$0.05$^d$ & 29.13 & 20.9\\
   $\kappa^1$ Cet & 1.03 & 0.95 & 9.3 & 0.65$\pm$0.05$^{e,d}$ & 28.79 & 9.14\\
   $\beta$ Com & 1.10 & 1.10 & 12.4 & 1.6$^{+0.9}_{-0.1}$ $^e$ & 28.21 & 9.13\\
   15 Sge & 1.01 & 1.10 & 13.5 & 1.9$^{+1.1}_{-0.9}$ $^f$ & 28.06 & 17.69\\
   18 Sco & 0.98 & 1.02 & 22.7 & 3.0$^{+0.2}_{-0.6}$ $^g$ & 26.8 & 13.9 \\
   Sun & 1.00 & 1.00 & 27.2 & 4.6 & $\approx$27 & 1 AU \\
   $\alpha$ Cen A & 1.10 & 1.22 & 30 & 5.5$^{+0.0}_{-0.8}$ $^e$ & 27.12 & 1.34 \\
   16 Cyg A & 1.00 & 1.16 & 35 & 7.0$\pm$0.3$^h$ & 26.89 & 21.1\\
\hline                                   %inserts single line
\end{tabular}
}
\end{minipage}
\end{table}	
	
	The sample for this study was selected by basing it off the original Sun in Time project \citep{gudel07,guinan09}. We omitted \(\beta\ Hyi\) as it has a radius of nearly twice that of the Sun, this gives it a much lower \emph{log(g)} which implies that it is no longer on the main sequence (MS). 
    47 Cas B was excluded from the study as it does not have very well constrained parameters such as mass, radius or rotation period \citep{gudel07}. It is also the secondary of a close binary system 47 Cas with an orbit of semi-major axis 1.32 AU \citep{gudel98}. 
    The stars treated here are all G0-5 type stars in the MS phase. The Sun is also included in our dataset. \Cref{table:sample} lists the most relevant quantities (mass, radius, rotation, age and X-ray luminosity) of these stars for this work. Studying these solar-analogues over a wide range of ages enables us to explore how the solar wind has evolved. Age ranges are included in \Cref{table:sample} and \Cref{fig:mdot-joined}. Since the stars have different methods of age determination, they have varying degrees of accuracy in their ages. We note, however, that this particular sample of stars is very well studied in the literature and their ages are relatively accurate. The preferred fit for our data is with rotation as it is more precise, but we also include the fit with age.

    \subsection{Temperature-Rotation Relation}
    \label{sec:temp-rot}
    Unfortunately, observations cannot constrain the values for base wind temperature and density, which are fundamental input parameters for our model. There has been significant research into constraining base temperature and density, usually by assuming they evolve with age or rotation or other stellar attributes (e.g. \citealt{holzwarth07,cranmersaar11}). Here, we use the X-ray-rotation relation presented in \Cref{fig:tcor} of this paper to observationally constrain the base wind temperature.
    Although the X-ray emission and wind acceleration are believed to originate at different locations (closed and open magnetic field regions, respectively), both phenomena are magnetic in nature and, therefore, it is expected that changes in closed regions would also affect changes in open regions. We assume that the temperature of the corona of solar type stars is related to the temperature at the base of the wind. The model we use is fitted to a piece-wise function around the value of 1.4 $\Omega_{\odot}$. The relation we find for $T_{\rm cor}$ was scaled down to correspond to observed solar wind temperatures near the base. As a result, a factor of 1.36 difference between coronal temperatures and base wind temperatures for all stars in our sample was found, shown in \Cref{table:massloss}.
            \begin{equation}
    \label{eq:rot-T}
    T_{0}\ (\Omega < 1.4\ \Omega_{\odot}) = 1.5\pm0.19 \left( \frac{\Omega_{\star}}{\Omega_{\odot}} \right) ^{1.2\pm0.54} \rm MK
    \end{equation}
    \begin{equation}
    \label{eq:rot-T2}
    T_{0}\ (\Omega > 1.4\ \Omega_{\odot}) = 1.98\pm0.21 \left( \frac{\Omega_{\star}}{\Omega_{\odot}} \right) ^{0.37\pm0.06} \rm MK
    \end{equation}
These relations are shown as solid lines in \Cref{fig:tcor}. The errors in the exponents arise from fitting. Note that a single fit to the coronal temperature data is also possible, but provides a larger $\chi^2$. In light of the recent works presented in Section 1.1 (e.g. \citealt{vansaders16,booth17,beck17}), we proceed with the broken power law fit throughout this paper. Appendix A shows the results one would have obtained, in the case a single power law had been adopted.

	\subsection{Density-Rotation Relation}
    \label{sec:den-rot}
    Currently, there is no available method to accurately define the density at the base of the wind, making it a difficult parameter to prescribe for simulations. Observations of stellar mass-loss rates would provide meaningful upper limits to the base density, but these are available for only a small sample of stars \citep{wood14}. \citet{ivanova03} find a relationship between rotation and coronal density (\Cref{eq:rot-rho}, also used by \citealt{holzwarth07,reville16}), from X-ray luminosity observations. We adopt this relationship for the density at the base of the wind for our simulations (\Cref{table:massloss}).
\begin{equation}
\label{eq:rot-rho}	
 n_0 = n_{\odot} \left( \frac{\Omega_{\star}}{\Omega_{\odot}} \right) ^ {0.6} .
\end{equation}
where n represents number density and is related to mass density by $n = \rho / \mu m_p$, where $\mu = 0.5$ is the mass fraction of a fully ionised hydrogen wind and $m_p$ is the proton mass.

	\subsection{Radio emissions from stellar winds}
	\label{sec:radio}
	One possible way of estimating wind densities (and mass-loss rates, $\dot{M}$) is by detecting these winds at radio wavelengths. The plasma that makes up stellar winds can emit in radio through the process of thermal bremsstrahlung from ionised plasma. This originates from the inner regions of the wind, where the density is highest \citep{panagia75,wright75,lim96,gudel02}. We can estimate the level of thermal radio emission from these winds using our model. According to previous studies by \citet{panagia75,wright75,gudel02,vidotto17b} the radio flux produced by a wind is:
    %\citet{panagia75} present a way to calculate this radio emission, newly derived in \cite{vidotto17}, given by \Cref{eq:Sv}. 
%
    \begin{align}
    \label{eq:Sv}
    S_\nu &= 10^{-29} A(\alpha) R^2_{\star} \left[ 5.624 \times 10^{-28} I(\alpha) n^2_0 R^2_{\star} \right]^{\frac{2}{2\alpha -1}} \nonumber \\
        &\ \ \times \left[ \frac{\nu}{10GHz} \right]^\beta \times \left[ \frac{T_0}{10^4 K} \right]^\lambda \times \left[ \frac{d}{1 kpc}\right]^{-2}\ \rm mJy,
    \end{align}
where the functions $I(\alpha)$ and $A(\alpha)$ are    
    \begin{equation}
    \label{eq:I}
    I(\alpha) = \int_0^{\pi /2} (sin\theta)^{2(\alpha-1)} d\theta,
    \end{equation}
    \begin{equation}
    \label{eq:A}
    A(\alpha) = 1 + 2\sum_{j=1}^{\infty} (-1)^{j+1} \frac{\tau_c^j}{j! j(2\alpha - 1) -2}.
    \end{equation}
The indices $\beta$ and $\lambda$ in \Cref{eq:Sv} are defined as    
    \begin{equation}
    \label{eq:betalambda}
    \beta = \frac{-4.2}{2\alpha -1} + 2, \hspace{1cm} \lambda = \frac{-2.7}{2\alpha-1} + 1, 
    \end{equation}    
with $\tau_c = 3$ and $\theta$ representing colatitude in radians. The wind density decay index, $\alpha$ is defined as
        \begin{equation}
        n = n_0 \left[ \frac{R_{\star}}{r} \right]^{-\alpha}.
        \end{equation}
The density decay index will eventually become $\alpha = 2$ as the wind reaches asymptotic terminal velocity. Since the radio emission originates near the base of the wind, the $\alpha$ parameter is likely to be greater than 2. For each star we found $\alpha$ by estimating the rate of density decline in the 1-5 $R_{\star}$ range. This range should account for the majority of ``stronger'' radio emission, as it is the densest region. 
It is important to note that the estimation of radio flux \Cref{eq:Sv} is based on an isothermal wind, whereas in our model the wind is a polytrope, allowing the temperature to vary as it expands. This approximation for radio flux should give a good indication of flux from these stars as emission is only estimated in the 1-5 $R_{\star}$ range, within which the isothermal approximation is adequate. 

The region where half of the emission occurs has a size
	\begin{equation}
	\label{eq:Rv}
	\frac{R_{\nu}}{R_{\star}} = \left[ 4.23\times10^{-27} I(\alpha) n_0^2 R_{\star} \right]^{\frac{1}{2\alpha-1}} \left[ \frac{\nu}{10GHz} \right]^{\frac{-2.1}{2\alpha-1}} \left[ \frac{T_0}{10^4 K} \right]^{\frac{-1.35}{2\alpha-1}}, 
	\end{equation}
which we refer to as the `radio photosphere' of the star. This is an important parameter as it illustrates how close to the star the emission will emanate and whether the wind is optically thin, as described by \citet{panagia75}.

%%% RESULTS IN THIS SECTION %%%
\section{Evolution of global properties of the Solar Wind}
    \label{sec:results_global}
    
    \subsection{Mass-loss rate}
    \begin{table}
\caption{Stellar wind properties for each of the simulated solar analogues in our sample. Values are displayed for base wind density, temperature and mass-loss rates (cf. \Cref{fig:mdot-joined}). The chosen values of $n_0$ and $T_0$ for the solar wind are such to reproduce observations (see text).}             
% 	title of Table
\begin{minipage}{\textwidth}
\label{table:massloss}      % is used to refer this table in the text
%\centering                          % used for centering table
%\resizebox{0.5\textwidth}{!}{
\begin{tabular}{l c c c c c}        % centered columns (4 columns)
\hline\hline                 % inserts double horizontal lines
Star & $n_0$ ($10^{8}$cm$^{-3}$) & $T_0$ (MK) & $\dot{M} (M_{\odot} yr^{-1})$\\    % table heading 	
\hline                        % inserts single 	horizontal line
   EK Dra & 8.8 & 4.7 & $1.4\times10^{-11}$ \\	
   HN Peg & 6.6 & 3.9 & $6.9\times10^{-12}$  \\
   $\chi^1$ Ori & 6.3 & 3.8 & $8.8\times10^{-12}$ \\
   $\pi^1$ UMa & 6.2 & 3.7 & $7.3\times10^{-12}$ \\
   BE Cet & 4.8 & 3.2 & $3.1\times10^{-12}$ \\   
   $\kappa^1$ Cet & 4.3 & 3.0 & $2.0\times10^{-12}$ \\
   $\beta$ Com & 3.6 & 2.7 & $1.9\times10^{-12}$ \\
   15 Sge & 3.4 & 2.6 & $2.1\times10^{-12}$ \\
   18 Sco & 2.5 & 1.9 & $2.8\times10^{-13}$ \\   
   Sun & 2.2 & 1.5 & $3.5\times10^{-14}$  \\
   $\alpha$ Cen A & 2.1 & 1.4 & $4.5\times10^{-14}$\\
   16 Cyg A & 1.9 & 1.1 & $8.1\times10^{-15}$ \\
\hline                                   %inserts single line
\end{tabular}
%}
\end{minipage}
\end{table}
    
\begin{figure*}
    \centering
    \includegraphics[width=\hsize]{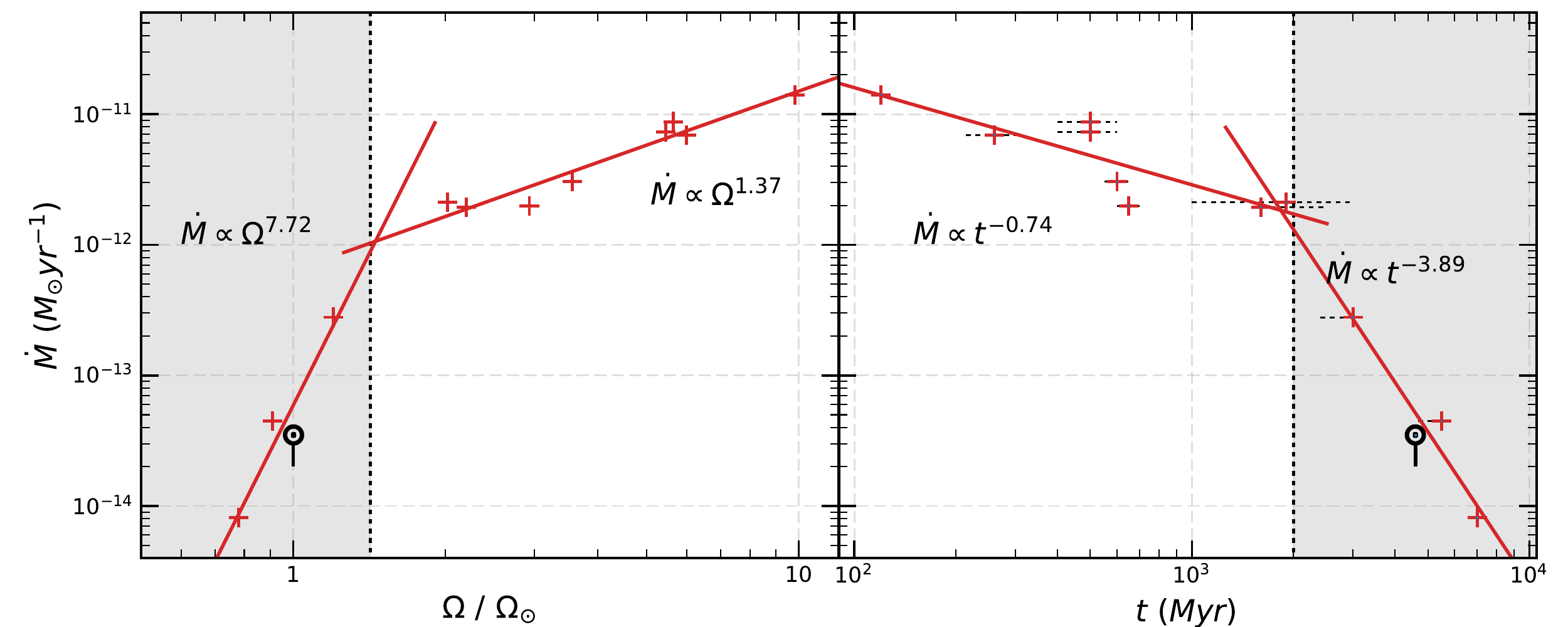}
      \caption{\emph{Left:} Calculated mass-loss rates (\emph{red crosses}) using our 1D polytropic model as the star spins down. \emph{Right:} Plot showing how the mass-loss rate of the Sun would change as it ages. Included in our calculation is our estimation of the current solar mass-loss rate ($\odot$) and the range of values calculated directly from observations (\emph{black solid line}) \citep{wang98}. Note the clear break in to a rapidly declining mass-loss rate regime (\emph{grey shaded region}) at $1.4 \Omega_{\odot}$ and $\approx$2 Gyr respectively. Errors in ages are shown as dotted black lines, with sources described in \Cref{table:sample}.
              }
         \label{fig:mdot-joined}
   \end{figure*}

    Since the model parameters in \Cref{table:massloss} are dependent on stellar rotation, we find in \Cref{fig:mdot-joined} that the mass-loss rate of stars is also dependent on rotation. The left panel of \Cref{fig:mdot-joined} shows a mass-loss rate that increases with rotation, with a break occurring at 1.4 $\Omega_{\odot}$. This dependence is as follows, 
    \begin{equation}
    \dot{M}\ (\Omega < 1.4\ \Omega_{\odot}) = 6.3\times10^{-14}\ \left(\frac{\Omega}{\Omega_{\odot}} \right)^{7.7 \pm 1.6} \rm M_{\odot} yr^{-1}
    \end{equation}
    \begin{equation}
    \dot{M}\ (\Omega > 1.4\ \Omega_{\odot}) = 6.3\times10^{-13}\ \left(\frac{\Omega}{\Omega_{\odot}} \right)^{1.4 \pm 0.15} \rm M_{\odot} yr^{-1}
    \end{equation}
    The mass-loss rate of solar-type stars is believed to decrease with time as the star ages. This is due to stellar spin down and a decrease in magnetic activity (e.g. \citealt{vidotto14}). Our results show a similar behaviour as the Sun evolves. However, our models predict a steep break in the mass-loss rate at an age of 2 Gyr (\Cref{fig:mdot-joined}), as a result of the break in $T_{\rm cor}$ with respect to rotation. The values of $\dot{M}$ we find for each star in our sample is shown in \Cref{table:massloss} and \Cref{fig:mdot-joined}, and follow the relations
    \begin{equation}
    \dot{M}\ (t \lesssim 2\ Gyr) = 5.0\times10^{-10}\ t_{\rm Myr}^{-0.74 \pm 0.19}\ \rm M_{\odot} yr^{-1}
    \end{equation}
    \begin{equation}
    \dot{M}\ (t \gtrsim 2\ Gyr) = 9.0\ t_{\rm Myr}^{-3.9 \pm 0.81}\ \rm M_{\odot} yr^{-1}
    \end{equation}
    where $t_{\rm Myr}$ is the age given in Myr. This shows, for example, that a young Sun of 100 Myr would have a mass-loss rate of \( 1.5\times10^{-11} M_{\odot} yr^{-1} \), almost 2.5 orders of magnitude larger than the current rate.

	\subsection{Radio emission from stellar winds}
	\begin{table}
	\caption{Radio properties of our wind models. Shown here are values for $\alpha$, which describes the behaviour of the density as a function of distance. $\beta$ and $\lambda$ describe the radio emission dependence on frequency and wind temperature respectively, which themselves depend on $\alpha$ (\Cref{eq:betalambda}). Outlined are results for the critical frequency, $\nu_{c}$, which describes the frequency below which the wind becomes optically thick. $S_{\rm vc}$ describes the radio flux at each respective critical frequency. }
    \label{table:Sv}
    \begin{tabular}{lccccc}
    \hline\hline
    Star & $\alpha$ & $\beta$ & $\lambda$ & $\nu_c$ (GHz) & S$_{\rm \nu c}$ ($\mu$Jy)  \\
    \hline
    EK Dra & 2.8 & 1.09 & 0.41 & 2.0 & 0.79 \\
    HN Peg & 3.1 & 1.19 & 0.48 & 1.7 & 1.9 \\
    $\chi^1$ Ori & 3.1 & 1.19 & 0.48 & 1.7 & 0.017\\
    $\pi^1$ UMa & 3.0 & 1.16 & 0.46 & 1.6 & 2.5 \\
    BE Cet & 3.3 & 1.25 & 0.52 & 1.4 & 0.70 \\
    $\kappa^1$ Cet & 3.4 & 1.28 & 0.53 & 1.3 & 2.5 \\
    $\beta$ Com & 3.4 & 1.28 & 0.53 & 1.2 & 2.9 \\
    15 Sge & 3.4 & 1.28 & 0.53 & 1.2 & 0.70 \\
    18 Sco & 4.0 & 1.40 & 0.61 & 0.99 & 0.48 \\
    $\alpha$ Cen A & 5.3 & 1.56 & 0.72 & 1.1 & 58  \\
    16 Cyg A & 6.4 & 1.64 & 0.77 & 1.0 & 0.16\\
	\hline 
	\end{tabular}
	\end{table}
    The densest parts of stellar winds might be able to produce free-free emission at radio wavelengths. Recently, \citet{fichtinger17} observed four stars, at 6 GHz and 14 GHz, using VLA and ALMA, namely: EK Dra, $\chi^1$ Ori, $\pi^1$ UMa and $\kappa^1$ Cet. Only two of these stars showed radio emission (EK Dra and $\chi^1$ Ori), however this emission did not emanate from their winds, but rather from the closed corona and flares. For $\pi^1$ UMa and $\kappa^1$ Cet no detections were made, which allowed the authors to place upper limits on the mass-loss rates of these winds (see also \citealt{vidotto17b}). \par 
    \Cref{table:Sv} shows the density fit parameters ($\alpha, \beta, \gamma$) we found for each star, with which we calculated radio emission over a range of frequencies (\Cref{eq:Sv,eq:I,eq:A,eq:betalambda}). We also computed the `radio photosphere' size (\Cref{eq:Rv}) for all the stars in our sample and found that all have their radio photosphere inside the radius of the star at both 6GHz and 14GHz. This implies that their winds are optically thin and do not emit at these frequencies. This agrees with the non-detections reported by \citet{fichtinger17}. To examine this further we computed the cut-off frequency, \(\nu_c \) (\Cref{table:Sv}), below which the radio photosphere surpasses the radius of the star and the wind becomes optically thick \citep{wright75}. From \Cref{eq:Rv}, the critical frequency below which these winds emit is given by
    \begin{equation}
    \label{eq:vc}
    \nu_c = [ 4.23\times10^{-27} I(\alpha) n_0^2 R_{\star}] ^{0.48} \left[ \frac{T_0}{10^4 K} \right]^{-0.64}\ \rm 10GHz .
    \end{equation}
    We calculate the expected flux density emitted from the wind at the same value of \(\nu_c\), given in \Cref{table:Sv} as \(S_{\rm \nu c}\). These flux densities are quite low with the exception of $\alpha$ Cen A as it is relatively close compared to the other stars. Note that, the wind cannot emit at frequencies larger than $\nu_c$. For stars in our sample, the cut-off frequency is around 1-2 GHz, implying that observations to detect these winds should be conducted at frequencies lower than 1 GHz. Also important to note is that, if the radio photosphere is very close to the surface of the star, any thermal emission is likely dominated by other stellar emission (i.e. coronal emission or flares). \par
    From \Cref{eq:vc}, we find that $\nu_c$ is weakly dependent on $\alpha$ and follows: $ \nu_c \propto n_0^{\rm 0.96} T_0^{-0.64}$. Since our model assumes that both base wind temperature and density rely on rotation, we can relate this cut-off frequency to rotation as    
    \begin{equation}
    \nu_c (\lesssim 1.4 \Omega_{\odot}) \propto \Omega^{-0.20}
    \label{eq:vc-omega2}.
    \end{equation}
    \begin{equation}
    \nu_c (\gtrsim 1.4 \Omega_{\odot}) \propto \Omega^{0.33}
    \label{eq:vc-omega1},
    \end{equation}
This means that there is an inflection in the dependence of $\nu_c$ with rotation. Although it is a weak dependence, it suggests that the lowest critical frequencies occur for stars at $\sim$1.4 $\Omega_{\odot}$.

	\section{Evolution of the local properties of the wind}
    
\begin{figure}
	\centering
    \includegraphics[width=\hsize]{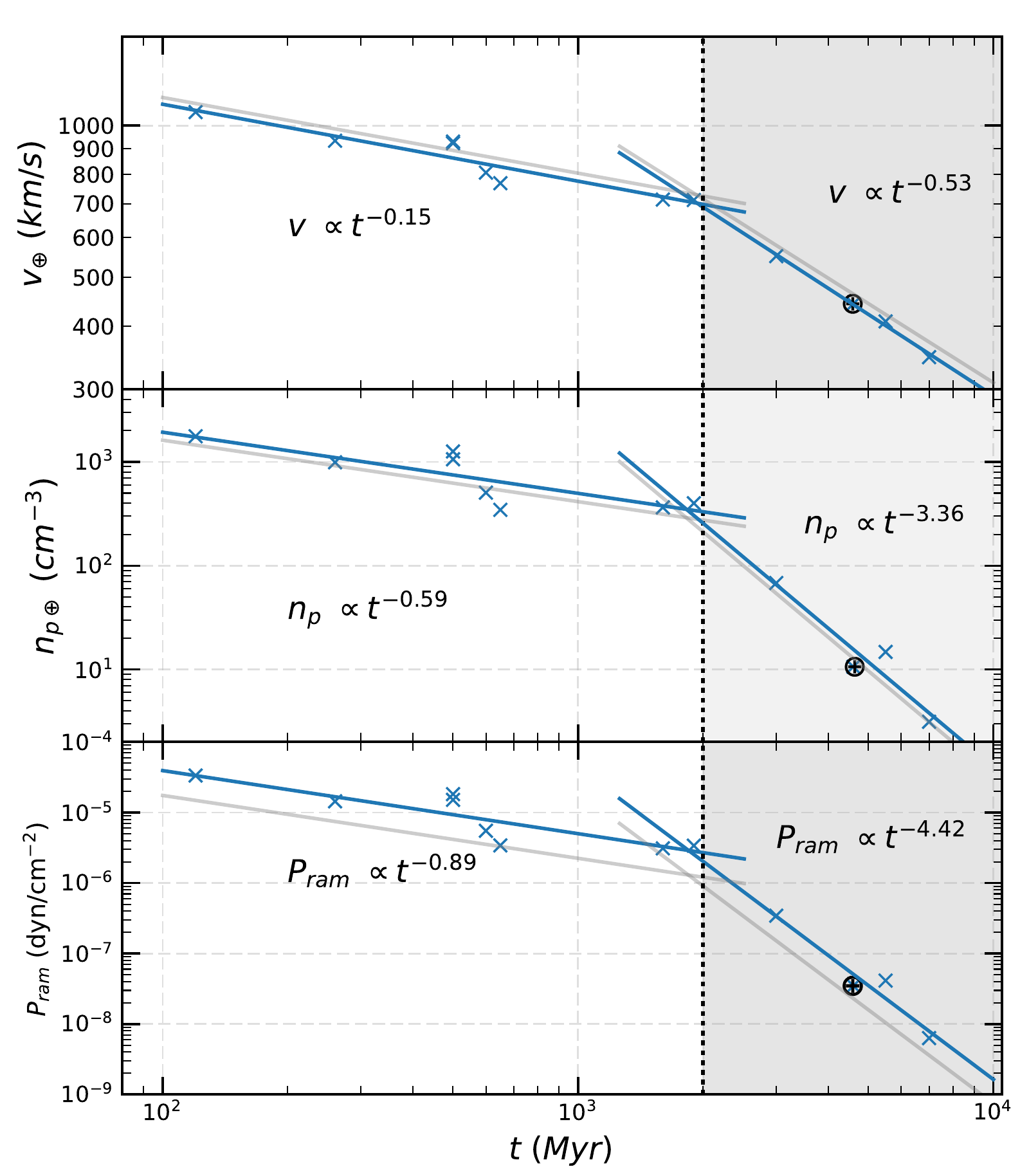}
    \caption{Local parameters for the solar wind in the vicinity of the Earth. We show local wind velocity (\emph{top}) and proton density (\emph{middle}) as stars evolve. This results in a present-day solar wind value ($\oplus$) at the Earth of 443 km/s and 10.5 $cm^{-3}$, respectively. From these values we calculate the expected ram pressure (\emph{bottom}) impinging on the Earth as it evolves. Shown are best fits to simulated values in separate regimes. Shown in grey are the expected values for a martian proxy planet orbiting each star.}
    \label{fig:vn-t}
	\end{figure}

    A direct output of our wind simulations is the local velocity and density for the wind at the position of the Earth. Therefore we can use these values to estimate the local velocity and density of the wind surrounding an evolving Earth as the system ages. The ram pressure that  impinges upon an evolving Earth is
    \begin{equation}
    	P_{ram} =  \rho_{\oplus} v_{\oplus}^2 .
    \label{eq:ram}
    \end{equation}
    This equation can be easily adopted to the martian case, by changing velocities and densities to those in the martian vicinity. In \Cref{fig:vn-t} we show the local wind velocity, $v_{\oplus}$, the local proton number density, $n_{\rm p\oplus}$, and the ram pressure, $P_{\rm ram}$ at the orbital distance of the Earth as the stars evolve (in blue; the grey line represents the martian values). Evidently, the break in stellar wind temperature, shown in \Cref{fig:tcor}, filters down to the local environment, which also displays a break in behaviour as these systems age. This happens at the age of 2 Gyr, denoted by a shaded region. This suggests that the young solar wind exhibited typical wind velocities of $10^3$ km/s at the orbital distance of the Earth.  \par 
    Once we know the ram pressure incident on the magnetosphere of the Earth, the magnetospheric standoff radius, R$_M$, at the sub-solar point can be calculated. This is done by balancing the ram pressure of the wind to the magnetic pressure of the planet's magnetosphere \citep{cravens97}.
    \begin{equation}
    	\frac{R_M}{R_{\oplus}}  = 1.4\left[ \frac{B_p^2}{8 \pi P_{ram}} \right] ^{\frac{1}{6}}.
    \label{eq:magsphere}
    \end{equation}
    $R_M$ is given here in planetary radii, \(B_p\) is the surface planetary magnetic field at the equator, which we assume to be the dipolar field only. The factor of $1.4$ accounts for currents that develop near the magnetopause boundary and produce their own magnetic fields \citep{cravens97,bagenal13}. \Cref{eq:magsphere} shows a dependence on stellar wind strength, which could have important ramifications for the development of life, as the solar wind is expected to have varied in the past on long timescales, as we have shown in \Cref{sec:results_global}. A smaller magnetosphere can lead to escalated atmospheric loss and reduced protection from the ionised wind (discussed further in \Cref{sec:discussion}). \par

	\begin{figure*}
	\centering
    \includegraphics[width=\hsize]{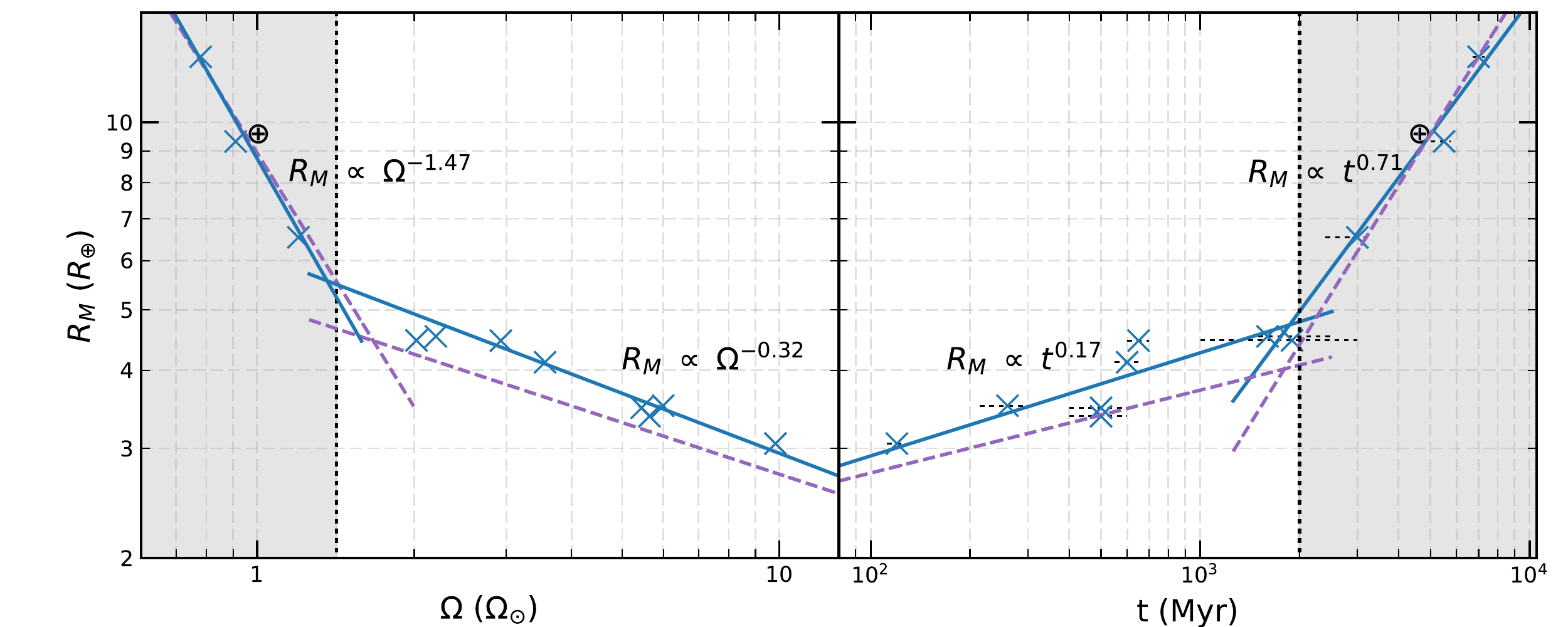}
    \caption{We observe trends in magnetospheric size with respect to rotation (\emph{left}) and age (\emph{right}). \emph{Blue x's:} Calculated standoff distances with a constant planetary magnetic field of 0.3G. The solar-Earth scenario as it is currently, given as \emph{$\oplus$}. \emph{Purple dashed line:} This depicts the standoff distances of the magnetosphere if the planetary magnetic field varied according to models described in \citet{zuluaga13}.}
    \label{fig:RM-joined}
	\end{figure*}

\Cref{fig:RM-joined} shows the trend in magnetospheric size for the Earth, both with solar rotation and age. We calculate the magnetospheric size considering the Earth's magnetic field to be constant with time (blue solid line) and varying with time according to the model by \citet{zuluaga13} (purple dotted line). This model is constrained by magnetic field measurements showing a 50\% field strength for a 1 Gyr system estimated by \citet{tarduno10}. We can see that both models differ only slightly, with the most significant differences occurring around 1-2 Gyr, showing a spread of just under 1 R$_M$. In particular, we find at early ages, both models predict magnetospheric sizes of $\approx$ 3 $R_E$. This is understandable, considering the weak dependence on the planetary magnetic field of 1/6, shown in \Cref{eq:magsphere}. We find a small magnetospheric standoff distance of 3.0 $R_E$ for a young Earth 100 Myr old, with the same planetary magnetic field as today ($B_p$ = 0.3 G). These standoff distances would be similar to extreme modern day events, such as the 2003 Halloween storm \citep{rosenqvist05}, caused by coronal mass ejections (CMEs). However, these transient CME events last only on the scale of hours, but at 100 Myr, these magnetosphere sizes would have been typical. Increased solar activity at younger ages (e.g. CMEs) would further compress the young magnetosphere, resulting in even smaller standoff distances. Younger stars are expected to be more active and would be more likely to produce large and more frequent transient events such as CMEs. 

\section{Discussion}
\label{sec:discussion}

	\subsection{Global Properties}
	%\label{sec:masslossdiscussion}
    Since direct observations of mass-loss rate and radio emission from stellar winds are difficult to obtain, we can use models to help understand the physical processes behind these winds. These models can provide information on the strength and location of the wind emission, therefore aiding in observing these winds directly. Recently, there has been research into the mass-loss rates of solar-analogue winds and how they would effect an orbiting planet, including a young Earth. While some work focuses directly on the Sun-Earth interaction \citep{sterenborg11}, others focus more on the stellar evolution of these types of stars \citep{wood02,wood05,wood14,cranmersaar11,matt12,fichtinger17} and their effects on exoplanets \citep{vidotto12,vidotto13,vidotto15,see14,see17,zuluaga16}.
    
    The main basis of our models is the temperature-rotation relation we presented in \Cref{fig:tcor}\footnote{\citet{holzwarth07} also derived a temperature-rotation relation which is based on activity-rotation relation. They attempted to constrain this dependence to a power law index, that lies somewhere between 0 and 0.5. They assumed a value of 0.1 for their model. This is a much weaker dependence than we find here for the slower rotators, but their ranges are within the values for the faster rotators.}. This type of relation between wind temperatures and X-ray observations, which has a strong correlation, is unique in predicting the winds of these stars. The break we find in wind temperature filters through to the trend in stellar mass-loss rate (\Cref{fig:mdot-joined}). We see a clear decay in $\dot{M}$ with stellar spin down and aging, with a break in $\dot{M}$ occurring at both 1.4 $\Omega_{\odot}$ and 2 Gyr respectively. Since our stellar wind models depend on the base wind temperature, it follows that mass-loss rate displays a similar trend. 
    
    Other models also predict a decay in mass-loss rate with stellar evolution. \citet{cranmersaar11} developed an Alfv\'{e}n-wave driven model for predicting winds from cool, late-type stars. They based their models on physically observed parameters from 47 stars of types G, K and M. \citet{johnstone15a,johnstone15b} employed a different approach and used polytropic wind models to reproduce the rotational evolution of solar-mass stars in open clusters. Both of the models found a mass-loss rate for young suns (at an age of 100 Myr) that are 1-2 orders of magnitude smaller than our predictions. However, for older stars, our predictions become smaller than theirs, since our model shows a steep decay in $\dot{M}$ for stars older than $\approx 2 $Gyr.
    Note that \citet{johnstone15b} assumed that the wind saturates for very young stars (<600 Myr) in the fast rotating track. When plotted in the $\Omega$-age diagram, the stars in our sample follow the 50th percentile track, as defined by \citet{gallet13}, which implies that they would not be part of the saturated regime explored by \citet{johnstone15b}. 
    
    Regarding the mass-loss rate with age, in our models, for ages younger than 2 Gyr, we found that $\dot{M} \propto t^{-0.74}$. This is much flatter than the original dependence derived by \citet{wood14} ($t^{-2.33}$), which has been revised as $t^{-1.46}$ by  \citet{johnstone15b}. In their Alfv\'{e}n wave-driven wind models, \citet{suzuki13} predicted $t^{-1.23}$, while \citet{cranmersaar11} predicted $t^{-1.1}$. Given the uncertainties in age measurements, our derived age-dependence is consistent with these works. Note also that, if we were to fit one single power law to a temperature-$\Omega$ relation, this would imply that the wind mass-loss rate would not have the change in regimes that we suggest, and the corresponding age dependence would be $t^{-1.36}$, a unique power law for all ages. 
   \citet{see17} investigated the trends in mass-loss rate with stellar age by adopting a potential field source surface model. By doing so they could investigate the topology of the coronal magnetic field of stars, including the extent of open flux regions which in turn allows angular momentum and mass-loss rates to be determined. They found lower mass-loss rates than presented here for all overlapping stars in our sample, but with a similar trend with age. 
   
    These mass-loss rates are important for solar and terrestrial evolution, as it affects solar evolutionary models, which in turn, directly effect the Earth through particle and radiation flux. A famous problem arose when these models predicted that the Sun would have been 25-30\% fainter than it is today \citep{newman77,gough81}, leading to a completely frozen Earth and Mars. Yet this prediction is inconsistent with the evidence suggesting that there was liquid water on the surface of both planets, implying planetary temperatures were not freezing \citep{sagan72}. This is called the faint young Sun paradox \citep{feulner12}. It leads us to the conclusion that there must be something awry with the standard solar model or the estimates surface temperatures of Earth and Mars. One solution to this issue is if the mass-loss rate of the Sun was higher in the past than expected. We integrated the mass-loss rate evolution calculated here and find that the Sun has lost 0.8\% $M_{\odot}$ since an age of 100 Myr. Our model results in a higher mass-loss rate at younger ages than previous models, while also predicting lower mass-loss rates at older ages. Even with the increased mass-loss rate from this model in the past, it does not solve the faint young Sun paradox, where a mass loss of 3-7\% $M_{\odot}$ is required.
    
	\subsection{Local Properties}
We have shown in \Cref{fig:RM-joined} that a young Earth orbiting a young Sun would possess a significantly smaller magnetosphere. In our results for magnetospheric standoff distance, we assumed that the magnetic dipole moment of the Earth has remained the same up to this day. However it is believed that this is not the case, although there is no consensus on how it has changed or by what amount. \citet{tarduno10} showed from ancient silicate crystals that the paleo-magnetic field of the Earth was much weaker than today, and estimated it to be 50-70\% of the current strength. We took a model which is constrained by these parameters and describes how the magnetic moment of the Earth has evolved over time from \citet{zuluaga13}. Using these values for $B_p$, we calculated $R_M$, shown as the purple dashed line in \Cref{fig:RM-joined}. This plot shows how significant the changes in the Earth's magnetic field can be over time, when calculating $R_M$. At 100 Myr we can see that the magnetosphere is approximately 2.8 $R_{\oplus}$ using a varying magnetic field, which is almost the same size derived with a constant magnetic field ($3\  R_{\oplus}$).
	\citet{see14} conducted a study into the effects that winds from solar type stars have on magnetosphere sizes of planets. They investigated how varying host star mass influenced the magnetospheric size and how this could effect habitability on any Earth-like exoplanets. This work complements the results of \citet{see14} as it shows how magnetosphere size will vary across different ages of solar analogues. Both \citet{vidotto13} (M dwarfs only) and \citet{doNasci16} ($\kappa^1$ Ceti) considered how stellar winds effected the interplanetary medium and how this could have impacted habitability on Earth-like exoplanets orbiting these stars. Also of interest is \citet{airapetian16}, who used 3D MHD Alfv\'{e}n wave driven models and found a paleo-solar wind that is twice as fast and 50 times as dense at 1AU at an age of 0.7 Gyr. We found very similar results for the wind of 1.9 times the velocity and 58 times as dense at 1AU for similar epochs.

	For a stronger solar wind, the `polar opening' region of the young Earth's magnetosphere, defined by the region which is covered by open magnetic field lines that extend into the magnetotail (in contrast to the closed dipolar magnetic regions in lower latitudes), can extend significantly further down in latitude than it does presently ($71.9^{\circ}$, \citealt{tarduno10}). We estimate that at an age of 100 Myr the polar opening region would extend as far as 55-60$^{\circ}$ from the equator. This larger polar opening region would have many implications for life developing on Earth, namely reduced protection from the harmful solar wind, and increased rates of atmospheric loss, through the expansion of the atmosphere and loss of volatiles. To intensify this effect, a younger Sun would be expected to be more active, with increased flaring rates and energy, potentially leading to additional atmospheric loss.

\Cref{fig:vn-t} also shows the local parameters of the wind around Mars. This allows us to calculate the height of the ionosphere, which acts to produce an induced magnetosphere above the martian surface. This is done by equating the stellar wind ram pressure with the thermal pressure of the martian atmosphere, taking today's values for the latter (7800 dyne/cm$^{-2}$, \citealt{harri14}). We find a present day value of 292 km, which is consistent with observations \citep{hanson77,withers09}. Assuming the current state of the martian magnetosphere, which is very weak given the lack of a global magnetic field, our models find the ionosphere height increases as the Sun-Mars system evolved, in a similar fashion to the Earth's magnetosphere size. It also predicts that for the immediate future, the ionosphere height will continue to grow. The change found in martian ionosphere height seems small compared to changes in Earth's magnetosphere over the same time period. We find a 42\% increase in ionosphere height for Mars from 100 Myr (218 km) to 7 Gyr (310 km). This is much smaller than the 324\% increase found in the Earth's magnetosphere from 100 Myr (3 $R_{\oplus}$) to 7 Gyr (12.7 $R_{\oplus}$). This arises due to the martian ionosphere size depending on the inverted natural logarithm of the wind ram pressure, giving a weaker dependence on the solar wind.

 \section{Conclusions}

	We simulated winds of a sample of solar analogue stars using 1D polytropic models. We selected our sample based of the `Sun in Time' program, which aimed to find a comprehensive trend in the evolution of solar activity \citep{ribas05,gudel07}. We presented a new rotation-temperature relation, that we used as an input for our simulations based off X-ray observations from \citet{johnstone15c}. We found a break in base wind temperatures at 1.4 $\Omega_{\odot}$. This leads to a sharp decline in wind temperatures as stars spin down past this point. We found stars rotating slower than this follow $T_{wind} \propto \Omega^{1.20}$, and stars rotating faster follow $T_{wind} \propto \Omega^{0.37}$. 
	
    This dependence between the wind temperature and rotation is rooted in the coronal temperature-rotation dependence. Although both the wind and the corona originate from different magnetic geometries (open and closed field lines, respectively), they are both caused by magnetism and therefore it is natural to assume that they will follow similar trends. The base temperature of the stellar wind is an important parameter for our simulations. It defines the rate of acceleration of the wind as it is launched from the surface of the star. Yet it is very difficult to constrain without direct observations from the stellar winds, which leaves only semi-empirical methods, which we employ here, to define the temperatures of these winds. 

	We found that the rate of mass loss from these stars seems to decline rapidly after 2 Gyr, with $\dot{M} = 8\times10^{-15} M_{\odot}yr^{-1}$ for a Sun of 7 Gyr. This steep decay in $\dot{M}$ could explain why older stars are inefficient at losing angular momentum, as shown by the atypically high rotation rates found in some older stars \citep{vansaders16}. 
    
    Our simulations provided us with the necessary parameters to make estimations on the thermal bremsstrahlung emissions from the winds of stars in our sample. We found that their winds only become optically thick below their critical frequencies, $\nu_c \approx 1-2$ GHz, which has a shallow dependence on stellar rotation as follows: $\nu_c(> 1.4\ \Omega_{\odot}) \propto \Omega^{0.33}$ and $\nu_c(< 1.4\ \Omega_{\odot}) \propto \Omega^{-0.20}$. We presented estimates for their fluxes at this critical frequency, $S_{\rm \nu c}$ (\Cref{table:Sv}). These values are pivotal to observing these stellar winds, and could explain some non-detections by previous attempts \citep{fichtinger17}.
    \Cref{eq:vc-omega1,eq:vc-omega2} show that there is an inflection in the rotation dependence of $\nu_c$, although the dependence is relatively shallow. Stars rotating faster than 1.4 $\Omega_{\odot}$ have higher cut-off frequencies.

We demonstrated the effects the aging solar wind has had on the evolving Earth, showing a steep increase in the growth of our magnetosphere since an age of 2 Gyr. We estimated the size of the magnetosphere at young ages to be {\raise.19ex\hbox{$\scriptstyle\sim$}}3 $R_{\oplus}$ at 100 Myr. This could have had implications for development of life due to the increased loss of atmosphere and a decrease in shielding ability from the solar wind. We found similar trends in the ionospheric height above the martian surface, yet the effect is not as extreme. 

	Although the young Sun's mass-loss rate had a shallow decline up to 2 Gyr, the total mass lost is still quite small. From our model we estimate a total mass loss of 0.8\%$M_{\odot}$, which is not enough to solve the faint young Sun paradox. This paradox has been studied at length, and a total mass loss of 3-7\% is required to solve it.

	Our model provides a semi-empirical method for determining base wind temperatures from X-ray observations of stars, which, in turn, allows an in depth analysis of the wind conditions surrounding these stars. Our current model did not allow us to evaluate angular momentum losses. This will be developed further by incorporating realistic distributions of stellar surface magnetism. The work here will be the initial foundation of a forthcoming 3D study into the winds of solar analogues.

\section*{Acknowledgements}
DOF acknowledges funding from a Trinity College Postgraduate Award through the School of Physics.
We also thank the Jorge Zuluaga for his constructive comments on the paper.

%%%%%%%%%%%%%%%%%%%%%%%%%%%%%%%%%%%%%%%%%%%%%%%%%%



%\bibliographystyle{mnras}

\bibliographystyle{mnras}
\bibliography{bibliography}

% %%%%%%%%%%%%%%%%%%%%%%%%%%%%%%%%%%%%%%%%%%%%%%%%%%

% %%%%%%%%%%%%%%%%% APPENDICES %%%%%%%%%%%%%%%%%%%%%

\appendix
\section{Tcor vs Omega: Goodness of fit}
\label{app:A}
In this work we present a fit between coronal temperature data and rotation rate for our sample. We find that a broken power law best describes the trend in data, although we note that there are other possible fits to the data. Having carried out reduced $\chi^2$ analysis on the goodness of fit, we find that both the low-$\Omega$ and high-$\Omega$ fits shown in \Cref{fig:tcor} have reduced $\chi^2$ of 3.5 and 1.4 respectively, while a single power law fit would have a reduced $\chi^2$ of 4.7. This would suggest that the broken power law fit produces a better result than a single power law. In addition, our choice of a broken power law is more in line with recent results (e.g. \citealt{vansaders16,metcalfe16,kitchatinov17,booth17,beck17}). We use this broken power law fit as an explanation as to why \citet{vansaders16} find anomalously high rotation rates in older stars; we propose that these lower coronal temperatures will lead to cooler winds, causing lower mass loss rates and therefore reduced angular momentum loss. Simultaneously, it also agrees with the break in X-ray luminosities found by \citet{booth17} in older stars.

\begin{figure}
    \centering
    \includegraphics[width=\hsize]{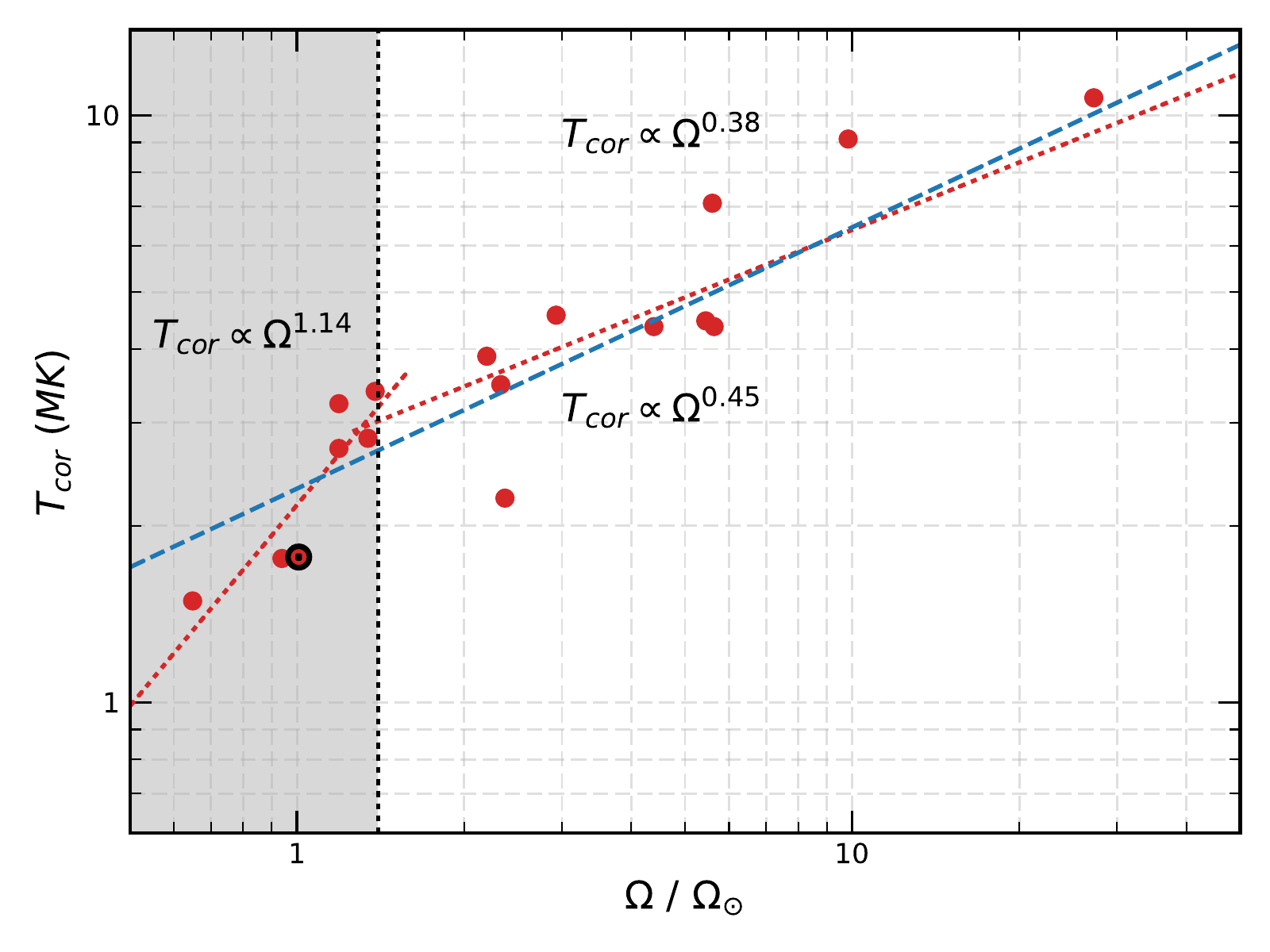}
      \caption{Comparison between a broken power law fit, which is used in this work, shown in red and a single power law fit, shown in blue.}
         \label{fig:onefit}
   \end{figure}

\begin{figure*}
    \centering
    \includegraphics[width=\hsize]{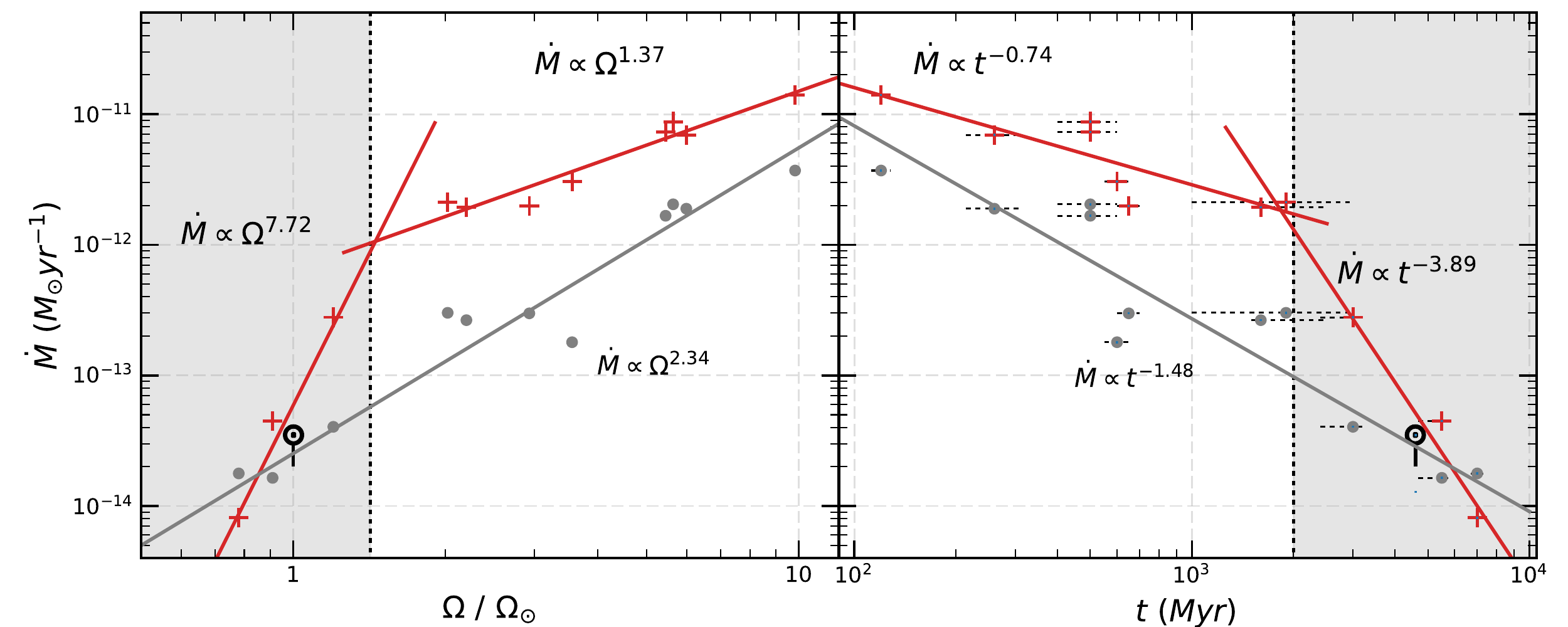}
      \caption{Plot similar to \Cref{fig:mdot-joined} with the resulting mass-loss rate from a linear fit in T$_{\rm cor}$-$\Omega$ (\Cref{fig:onefit}) shown in grey. Other symbols represent the same as in \Cref{fig:mdot-joined}. We can see this produces lower mass-loss rates in most cases, with the exception of 16Cyg A. The relation for each fit is shown beside the fitted line.
              }
         \label{fig:mdot-singlefit}
   \end{figure*}

For completeness we calculate the mass-loss rates that would result from a single power law fit. Shown in blue in \Cref{fig:onefit} is the fit produced by using a single power law to fit the coronal temperatures with stellar rotation. We find a relationship of $T_{\rm cor} \propto \Omega^{\rm 0.45}$ for a single power law. This value lies between both values found for the broken power law (1.14 and 0.38). A single power law results in much higher temperatures as you move to slower rotators. We can see how this fit affects the mass-loss rates in \Cref{fig:mdot-singlefit}. We find that $\dot{M} \propto \Omega^{2.34}$ and $\dot{M} \propto t^{-1.48}$. This fit produces lower mass-loss rates than our previous broken fit for any stars younger than the crossing point of both methods ($\approx$ 5500 Myr or 8.5$\Omega_{\odot}$). It also results in higher mass loss rates for stars older than 5500 Myr, in our case only showing as an increased mass-loss rate for 16 Cyg A. If this new $\dot{M}$-Age relationship is integrated from 100 Myr to the present (4600 Myr), we find a total mass lost of 0.14\% the current solar mass.
%
% %If you want to present additional material which would interrupt the flow of the main paper,
% %it can be placed in an Appendix which appears after the list of references.

% %%%%%%%%%%%%%%%%%%%%%%%%%%%%%%%%%%%%%%%%%%%%%%%%%%

% Don't change these lines
\bsp	% typesetting comment
\label{lastpage}
\end{document}